\documentclass[acmsmall]{acmart} 
\AtBeginDocument{%
  }

\setcopyright{cc}
\setcctype{by}
\acmJournal{PACMHCI}
\acmYear{2026} \acmVolume{10} \acmNumber{5} \acmArticle{MHCI3557}
\acmMonth{8} \acmDOI{10.1145/3821659}

\usepackage{siunitx}
\usepackage{enumitem}
\usepackage{xcolor}
\usepackage{soul}

\begin{document}
\title[Intention-Behavior Gap in Smartphone Usage]{To Stop or Not to Stop: Exploring the Intention-Behavior Gaps in Smartphone Usage}


\author{Jian Zheng}
\email{jzheng23@umd.edu}
\orcid{0000-0002-6075-9313}

\author{Eun Kyoung Choe}
\email{choe@umd.edu}
\orcid{0000-0001-5038-8320}
\affiliation{%
  \institution{University of Maryland}
  \city{College Park}
  \state{Maryland}
  \country{USA}
}

\renewcommand{\shortauthors}{Zheng et al.}

\begin{abstract}
As smartphones become integral to daily life, researchers have sought to identify when the use becomes problematic.
Previous studies have operationalized problematic smartphone usage (PSU) from either an intention or a behavior perspective. Both risk delivering interventions not welcomed by users. 
We propose a novel approach to operationalizing PSU as the intention–behavior gap (IBG). We collected self-reported data on intentions to stop phone usage, alongside usage behavior data, from 37 participants over two weeks. We calculated IBG, examined effects of demographic and contextual variables, and developed machine learning models to predict IBG in real time.
We found that IBG was explained by gender, time, app, and input interactions, among other factors. 
Intention was predicted most accurately with only personal data, whereas behavior and IBG were predicted most accurately with both personal and global data.
Our findings can inform the design of future intervention tools optimized for timing and adaptive intensity. 

\end{abstract}




\maketitle

\section{Introduction}

From 2019 to 2024, the number of smartphone users worldwide has increased from 2.27 billion to 4.88 billion, and the penetration rate has increased from 34\% to 60\% \cite{Gill2025-bg}. This rapid growth highlights the increasing reliance on smartphones in modern society. People use smartphones for both productivity and entertainment. However, they often use smartphones in problematic ways: they check their phones too frequently, use phones for too long, or at inappropriate times (e.g., while driving or in the dark). Such problematic smartphone usage (PSU) can cause physical, psychological, and social consequences \cite{Busch2021-fi}. 

To assist people in controlling their phone usage, researchers and practitioners have developed various intervention tools \cite{Biedermann2021-rj, Lyngs2022-xg, Monge-Roffarello2022-mw, Nwagu2026-ft}. Interventions are typically triggered by users' specific behaviors, such as starting to use certain apps \cite[e.g.,][]{Kim2019-sk, Wu2024-xe}. This trigger mechanism is straightforward but lacks flexibility. Blocking a selected app can disrupt normal usage if users feel a genuine need to use it. 
An alternative approach involves predicting users' intention to stop using machine learning models and using this prediction to trigger interventions \cite{Orzikulova2024-yd}. This intention-based trigger, however, can result in unnecessary interventions if users can stop by themselves. Even worse, telling users to stop when they can do it independently may threaten their perceived autonomy and cause psychological reactance \cite{Brehm2012-nv}.

We propose to operationalize PSU as the intention-behavior gap (IBG) \cite{Conner2022-zv}. Smartphone usage becomes problematic when there is a gap between the intention to stop and the behavior of continuing. In other words, usage is problematic when users want to stop but are unable to do so by themselves. In such ``should but cannot stop'' cases, interventions are necessary and more likely to be appreciated than those triggered by either behavior or intention alone. Following previous studies on PSU, we explore the effects of demographic and contextual factors as potential antecedents of IBG. Finally, although this study does not contain an intervention, we explored the feasibility of predicting IBG in real time through machine learning to inform the design of future intervention tools.

We aim to answer three research questions (RQs):
\begin{itemize}
    \item RQ1: How to operationalize problematic smartphone usage as the intention-behavior gap?
    \item RQ2: What factors, both demographic and contextual, lead to larger intention-behavior gaps in smartphone usage?
    \item RQ3: How to predict the smartphone usage intention-behavior gap in real time?
\end{itemize}

We collected both smartphone usage intentions via the experience sampling method (ESM) and behavior data from 37 Android users over at least two weeks. Subsequently, we interviewed 19 of them about the gaps we observed in their data. We demonstrated how to operationalize PSU as IBG. Our findings indicate that IBG is influenced by factors such as gender, time, app in use, input interaction, physical activities, and session length. Intention is most accurately predicted by the personal model, whereas behavior and IBG are most accurately predicted by the combined personal-global model.

Our primary contribution is the operationalization of PSU using the concept of IBG, along with insights into demographic and contextual factors that influence IBG. A secondary contribution is the prediction of PSU as a continuous variable using machine learning, offering benchmarks for future research and informing the development of intervention tools. A third contribution is a dataset of 5,772 phone usage sessions containing intention, behavior, and contextual factors from 37 users with their demographics, available on the Open Science Framework (OSF).


\section{Related Work}

\subsection{Problematic Smartphone Usage}
Smartphone usage can become problematic in a variety of ways. Common forms include infinite scrolling on social media \cite{Rixen2023-wg}, binge-watching of short videos \cite{Lu2022-sr}, and habitual checking the phone \cite{Oulasvirta2012-en}. Certain contexts could render nearly any phone usage problematic, such as use in bed during sleep hours \cite{Bernroider2014-ye}, in classrooms \cite{Rozgonjuk2018-bf}, while driving \cite{Soror2012-us} or walking \cite{Tao2016-sk}, and during face-to-face interaction (a behavior referred to as phubbing) \cite{Chotpitayasunondh2018-ty}. 

\subsubsection{Operationalization of Problematic Smartphone Usage}
Although both researchers and lay users have a general understanding of PSU, the definitions are still evolving \cite{Busch2021-fi}. Researchers have operationalized and thus measured PSU in diverse ways.  
Some researchers have operationalized PSU on the individual level. They measure the level of PSU with questionnaires, such as the Smartphone Addiction Scale \cite{Kwon2013-ur, Kwon2013-zm}, to categorize users. This approach can investigate who is more prone to PSU and reveal the trend across years, countries, and demographic groups \cite[e.g.,][]{Olson2022-mg, Shin2013-lh}. However, the results can provide little guidance on intervention. Having a high PSU score does not mean all the person's phone usage is problematic. 

To distinguish problematic from non-problematic usage rather than categorizing users, researchers have operationalized PSU at the usage session and app usage episode levels, which are defined as follows: ``usage sessions are continuous, uninterrupted sequences of device use that occur between unlocking and locking the device,'' and ``application usage episodes are continuous, uninterrupted sequences of app use that occur between launching and closing an app'' \cite[][p. 9]{Parry2025-io}. 
ESM was often used to inquire about users' subjective feelings to label problematic usage \cite{Lukoff2018-ws, Tran2019-mw}. The problematic nature of each usage instance is determined by a single self-report from the user, typically collected at the start of the session or episode. 
While these studies may not use the term PSU, they all characterize a subset of phone usage as undesirable, in contrast with usage that is considered normal, acceptable, or justified. 

Recently, researchers have examined within-session variations at the intra-session level. For example, Finesse \cite{Cho2021-xe} asked users to select which features within an app (e.g., following posts or suggested posts on Instagram) they regretted using.  
Time2Stop \cite{Orzikulova2024-yd} prompted users at app launch and exit and every ten minutes during use to indicate whether it would be better not to use that app at the moment. Across individual, inter-session, and intra-session levels, PSU has always been measured through users' self-report. Interventions, however, follow a different pattern. 
\subsubsection{Intervention against Problematic Smartphone Usage}
To assist users in controlling their phone usage, researchers and practitioners have developed various intervention tools \cite{Biedermann2021-rj, Lyngs2022-xg, Monge-Roffarello2022-mw, Nwagu2026-ft}. 
We see a mismatch between how PSU is measured and how it is intervened with. PSU is measured using subjective, self-reported data collected through one-time surveys or ESM. Interventions, on the other hand, are usually rule-based and triggered by objective behavioral signals, such as unlocking the phone screen or launching specific apps \cite[e.g.,][]{Kim2019-sk, Wu2024-xe}, or using them for a predefined duration \cite[e.g.,][]{Terzimehic2024-ji, Lu2024-dc}. For example, InteractOut \cite{Lu2024-dc} inhibits the natural execution of gestures (e.g., tap twice, instead of once, on an icon to open an app) after one cumulative hour of use across monitored apps. 

Those behavioral triggers are straightforward but do not adapt to users' intentions. First, long screen time alone does not predict negative well-being \cite{Katevas2018-nm}. Intervention targeting all phone usage may impede users' work-related tasks. Second, restrictions on certain apps could disturb and annoy users if they have justified reasons. Even the most entertainment-oriented apps (e.g., TikTok) can be used for justified reasons (e.g., TikTok STEM feed). Users could feel they deserve some entertainment, for example, after working for a long time. Third, limits on screen time have an uneven effect through the day---users are more likely to get interventions in the evening, when they have run out of the time limit, than in the morning, when the time limit is just reset. To sum up, interventions will not work well when the users have the intention and reasons to use phones. 

Perhaps no one understands users' intentions better than themselves. However, repeatedly asking users whether they should stop is impractical. One solution is to first collect users' intention data and then predict future intentions through machine learning. For example, Time2Stop \cite{Orzikulova2024-yd} built a machine learning model to predict whether users would agree they should stop using their phones, intervening only when the prediction was ``agree.'' However, such a purely intention-triggered approach also has limitations: if the users want to stop and are able to do so on their own, delivering an intervention may trigger \textit{psychological reactance} \cite{Brehm2012-nv} and cause negative consequences. 

Psychological reactance is an unpleasant motivational response that arises when people's freedom to make personal decisions is threatened \cite{Brehm2012-nv}. It occurs with digital interventions \cite[e.g.,][]{Hu2025-dm, Tatum2018-mt}. For example, TikTok has introduced usage reminders to reduce excessive usage. However, those reminders threatened users' sense of freedom and caused psychological reactance \cite{Hu2025-dm}. Similarly, if phone users have the intention to stop and can actually stop by themselves shortly after, interventions could threaten their autonomy and cause psychological reactance. To prevent this from happening, we propose to build intervention triggers on the gap between intention and behavior. 

\subsection{Intention-Behavior Gap}
Intention does not always translate into the actual behavior---there could be a gap in between \cite{Webb2006-gq, Sheeran2018-pv}. 
The intention-behavior gap (IBG) has been mainly studied in the field of physical activity \cite[e.g.,][]{Pfeffer2017-wc, Pfeffer2020-sm, Sheeran2018-pv}, with other domains including medication adherence \cite{Faries2019-zw} and climate change behaviors \cite{Sinclair2025-wk}. 
To the best of our knowledge, no research has examined the IBG in the context of smartphone usage or more general digital well-being. Moreover, those existing IBG studies typically rely on self-reported behavior, but self-reported smartphone usage does not reflect actual behavior accurately \cite{Elhai2021-tw, Shaw2020-lb, Parry2021-bv}. In addition, previous research has examined IBG over weeks or months, whereas smartphone usage fluctuates across and even within sessions and needs to be studied with much finer granularity. 

The intention-behavior relationship can be decomposed into a 2 $\times$ 2 matrix that distinguishes between intending to act vs. not intending to act and subsequently acting vs. not acting. This framework yields four groups of individuals, corresponding to the four cells of the matrix: inclined actors, who intend to act and actually act; inclined abstainers, who intend to act but do not act; disinclined actors, who intend not to act but nevertheless act; and disinclined abstainers, who intend not to act and do not act. Among these groups, inclined abstainers are primarily responsible for the IBG \cite{Orbell1998-sh, Sheeran2002-vu}. We apply this intention-behavior matrix to smartphone usage, not at the individual level but at the usage session level. We focus on inclined abstainer sessions, or the overuse cases, where users want to stop but fail to stop using their phones. 
To build intervention tools targeting those sessions in real time, we need to predict the intention and behavior. 

\subsection{Predicting Smartphone Usage}
To include users' intention as a criterion of intervention trigger, we need to rely on machine learning to predict users' intention. The usage behavior we care about is the remaining session length, which is unknown in real time and also needs to be predicted. In this subsection, we summarize related work on the prediction of smartphone usage intention and behaviors.

Besides Time2Stop \cite{Orzikulova2024-yd}, there are several other studies trying to predict self-reported intention of phone usage. Those studies first collected users' subjective rating of whether their phone usage is problematic (e.g., time-killing \cite{Chen2023-dc, Fang2024-wv}, ritualistic vs. instrumental \cite{Hiniker2016-sd}) through ESM and behavior data through phone event log data \cite{Hiniker2016-sd}, screenshots \cite{Chen2023-dc}, or screen text \cite{Fang2024-wv}. Then they trained machine learning models to predict their subjective rating from the behavior data. 

Regarding smartphone usage behavior, researchers have been working on the prediction of the next application the user is going to open \cite{Baeza-Yates2015-se, Yan2012-lz, Huang2012-oh, Xia2020-nu, Shen2019-um}. For example, DeepApp \cite{Xia2020-nu}, a multitask deep learning framework, leveraged spatio-temporal and user-specific contexts to predict whether an app will be used on the next day. 
Another thread of research has employed users' typing behavior to predict the presence vs. absence of typing errors \cite{Simou2023-wp} or to detect emotion states \cite{Ghosh2017-fu}. There has been no research to date aiming to predict smartphone usage session length in real time nor to simultaneously predict both usage intention and behavior. 

In these research, the intention to be predicted is usually binary (e.g., ``agree'' vs. ``disagree'' to stop \cite{Orzikulova2024-yd}, time-killing or not \cite{Chen2023-dc, Fang2024-wv}, ritualistic vs. instrumental \cite{Hiniker2016-sd}, or ``problematic'' users vs. normal users \cite{Shin2013-lh}). As a result, the interventions work in an all-or-nothing way.
We believe it is better to predict PSU as continuous, consistent with its measurement, so that we can design interventions with varying levels of intensity. 

\section{Methods}
We built AwareMind, an Android app that collects smartphone usage intentions via ESM and behavior data. We then deployed it in a study with 37 Android users over at least two weeks and then interviewed 19 of them about the gaps we observed in their data.

\subsection{Design of AwareMind}

AwareMind is designed as a data collection tool. The release version is available on Google Play\footnote{\url{https://play.google.com/store/apps/details?id=com.jianzheng.studyzero}}, and the source code is available on GitHub\footnote{\url{https://github.com/jzheng23/AwareMind}}. It records the time zone information and the device information (i.e., Android version, brand, and model) once and only after the user logs in. It collects data on phone usage intention and behavior continuously through the data collection period. 

\subsubsection{Intention Data Collection}
\begin{figure}
  \centering
  \includegraphics[width=0.6\textwidth]{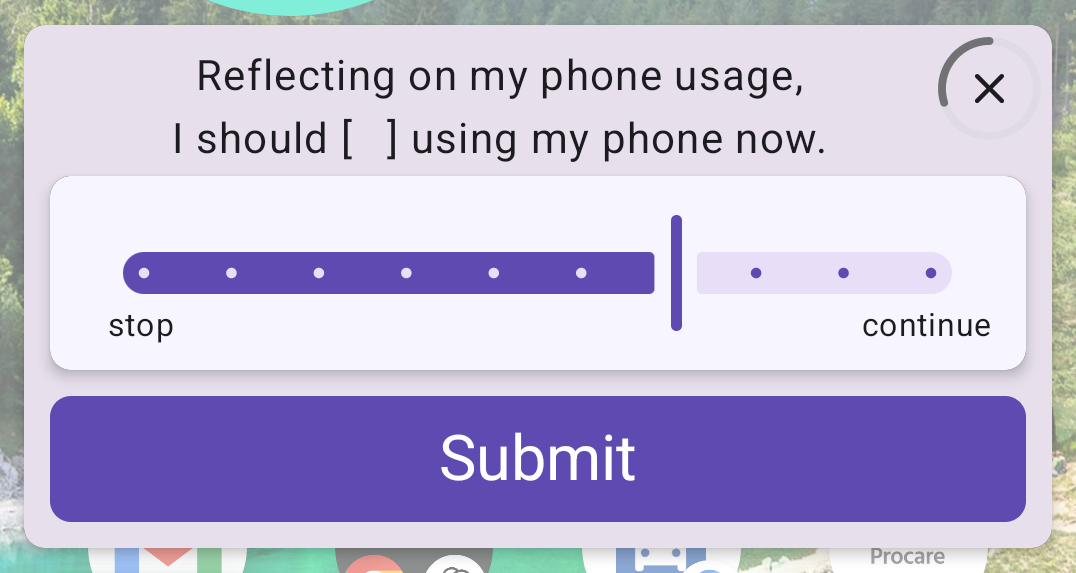}
  \caption{A screenshot of the ESM survey. The dismiss button and the auto-dismiss countdown appear in the top-right corner.}
  \label{fig:esm}
  \Description{A dialog box showing a reflection prompt with a horizontal slider. The slider has 10 discrete points, with the left end labeled "stop" and the right end labeled "continue." The current selection is near the middle. A purple Submit button appears below the slider, and an X dismiss button with circular countdown timer is in the top-right corner.}
\end{figure}

We collected the intention data using ESM.
During phone usage, AwareMind sends one-question surveys in an overlay display. The question reads, ``Reflecting on my phone usage, I should [ ] using my phone now,'' where participants fill in the blank using a 1--10 scale (1 = \textit{stop}, 10 = \textit{continue}); see Fig. \ref{fig:esm}. 
Single-item Likert scales have been used in previous studies with high sampling frequency, for example, to measure meaningfulness \cite{Lukoff2018-ws} and sense of agency \cite{Zhang2022-rm}. Time2Stop \cite{Orzikulova2024-yd} also used a single question (i.e., whether users should stop using the phone or not) to collect the PSU label.
AwareMind is configured to collect up to 16 responses per day. Participants can dismiss the survey. If unanswered, the survey is auto-dismissed after 30 seconds.
The surveys are designed not to appear while participants are driving.

\subsubsection{Behavior Data Collection}

AwareMind records users' phone usage behavior, including locking and unlocking of the phone screen, input interactions, and app usage. 
The input interaction data to be recorded are three types of interaction: taps, scrolls, and text edits. This feature relies on the Accessibility Service API.
It records the interaction type and the timestamp of each interaction. For scrolls, it records the distance of the scroll, both horizontally and vertically. For text edits, it does not record the content of input, but only the number of characters typed.
The app usage data to be recorded include the app package name, class name, starting time, and ending time. It does not record the app content. 

\subsection{Procedure}
After signing the consent form, participants installed AwareMind from Google Play, logged in with their assigned user ID, and granted the necessary permissions. Next, participants completed a brief tutorial where they learned how to respond to the surveys, and then they could start data collection. Over the following two weeks, they received multiple surveys per day. After the two weeks, those who had made 200 or more reports were told to stop the ESM data collection. Participants who had not met this criterion were offered two options: either to stop the ESM at that point or to continue for up to one more week in order to qualify for higher compensation.

\begin{figure}
  \centering
  \includegraphics[width=\linewidth]{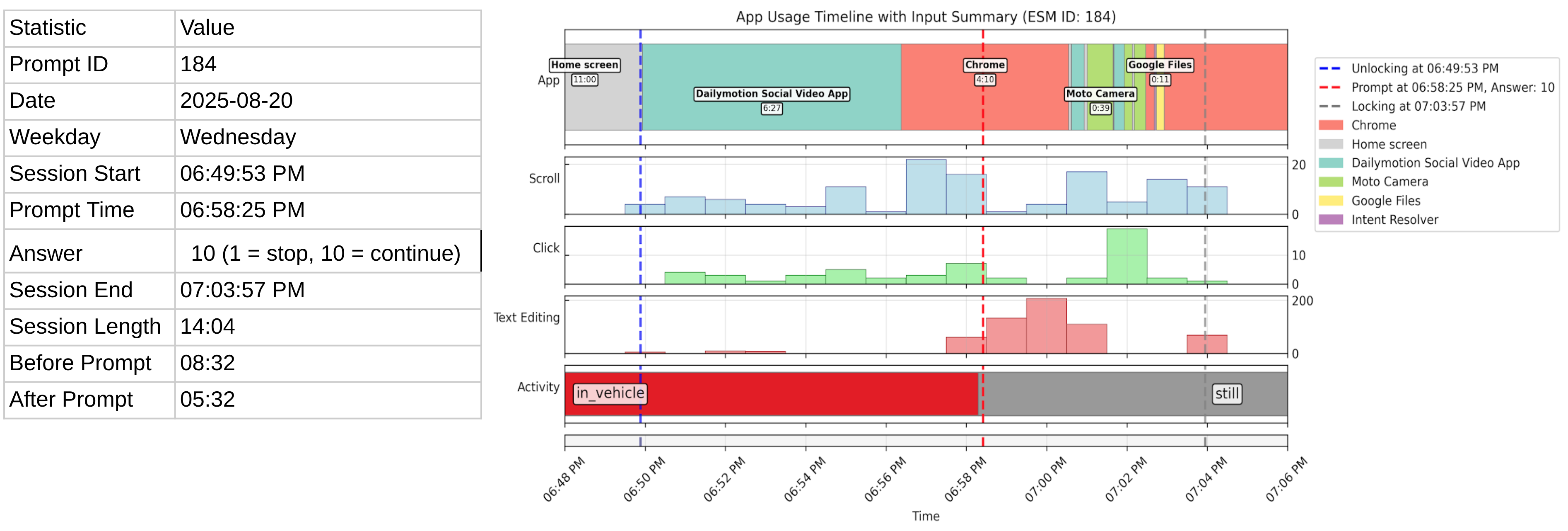}
  \caption{A screenshot of the visualization used in the interview. A table presents the key information of the usage session. A composite plot shows the app usage history on the top; in the middle it shows the input interaction (scroll, click, and text edit) count in each minute during the usage session; at the bottom it shows the detected physical activities.}
  \label{fig:interview}
  \Description{Left: Table with session metadata including date, times, prompt response (10 = continue), and session duration. Right: Composite timeline with app usage sequence (Home screen, DailyMotion, Chrome, Moto Camera, Google Files), per-minute bar charts for scroll, click, and text editing events, and physical activity (in vehicle to still). Vertical lines mark session start, prompt time, and session end.}
\end{figure}

After the ESM data collection, we randomly selected and invited 19 participants to a \textit{data engagement interview} \cite{Moore2021-rg}, from participants who met two criteria: (1) they completed more than 105 reports (i.e., a 50\% response rate), and (2) fewer than 95\% of their responses were 9 or 10. Each interview lasted about 30 minutes and was conducted via Zoom. The first author showed each participant about six cases where their actual behavior diverged from their reported intention (see Fig. \ref{fig:interview} for an example). These included instances where participants reported an intention to stop but did not do so promptly, as well as cases where they reported an intention to continue but stopped soon after.
To help participants recall the selected sessions, we reconstructed each usage session to discuss with a table and a composite plot. The table presents the information of the usage session. 
The composite plot has five panels. The top panel displays the app usage history, the three panels in the middle show the number of input interactions (scrolls, clicks, and text edits) per minute during the session, and the bottom panel illustrates the detected physical activities.
The first author asked participants to describe what had happened during the session and to explain the reasons for their choices and behavior. See Appendix for the interview protocol. All interviews were audio-recorded and automatically transcribed via Zoom. The first author checked and manually edited the transcriptions afterward. The study was approved by the Institutional Review Board (IRB) of the University of Maryland, College Park.  

\subsection{Participants}

We aimed for 40 participants according to the local standard \cite{Caine2016-hw}.\footnote{We checked twelve papers on PSU involving ESM since 2020 \cite{Orzikulova2024-yd,Rixen2023-wg,Terzimehic2023-yi, Zhang2022-rm, Terzimehic2022-si, Haliburton2022-ch, Cho2021-xe, Kim2020-it, Lukoff2018-ws, Katevas2018-nm, Giunchiglia2017-wg, Hiniker2016-sd} and found that the median sample size is 33.5 and the range is 10--340.}
We recruited participants through mailing lists, social media, undergraduate courses from the authors' institution, and snowball sampling. Recruitment criteria were being at least 18 years old, residing in the US, using an Android phone as their primary device, spending at least three hours per day on their phones, being interested in reducing excessive phone usage, and either habitually locking the phone screen when not in use or having the auto-lock function enabled.

Initially, 48 participants were recruited.
During the data collection, six dropped out due to technical issues. Five more were excluded during data analysis because they answered 10 (i.e., \textit{continue}) to over 95\% of all the ESM questions, which we saw as contradictory to their claimed interest in reducing excessive phone usage. 
The final sample consisted of 37 participants (22 female, 15 male). Participants ranged in age from 18 to 64 years ($M = 42.1, SD = 13.3$). The majority of participants identified as white ($n = 15$), followed by Black ($n = 14$), Asian ($n = 7$), and others ($n = 1$). Nine participants held graduate or professional degrees, nine held bachelor's degrees, eight had some college education without a degree, six held associate or technical degrees, and five had a high school diploma or GED. About half of all the participants were employed full-time ($n = 19$), while others had various part-time jobs, were students, were self-employed, or worked as homemakers or volunteers. Participants used their phones for about five hours each day on average during the data collection. After completing data collection through AwareMind, each participant was compensated with up to 30 USD ($M = 26$ USD, ranging from 13 to 30 USD) based on the number of reports they had submitted. Those who completed the interview ($n = 19$) were compensated with an extra 10 USD. 

\subsection{Data Analysis}
We used Python 3.12 to analyze the smartphone usage intention and behavior data. We care most about three variables: participants' self-reported intention to continue (later referred to as the \textit{intention}), their behavior of continuing measured as the remaining session length after answering the ESM questions (later referred to as the \textit{behavior}), and the intention-behavior gap calculated as the difference between behavior and intention (later referred to as the \textit{IBG}). We measure both the intention and the behavior as continuous variables. We measure intention on a Likert scale, which is an ``imperfect'' interval scale \cite{Borgatta1980-ym} but often treated as continuous (interval) \cite{Wu2017-yd}. We measure behavior as the remaining session length, which is a real continuous (ratio) variable.  

Although we included ``having the phone screen locked (either automatically or manually) when not in use'' as a recruitment criterion, we still found some \textit{inactive sessions}, where the phone was left unlocked, sometimes being charged, but not in active use for a long time. We detected and excluded those inactive sessions (202 sessions, fewer than 1\% of all sessions) with the app usage and input interaction data. 

We use QualCoder\footnote{\url{https://qualcoder.wordpress.com/}} to analyze the interview transcriptions. We focused on what the participants explicitly said more than the implicit meaning. We omitted some filter words (e.g., ``you know,'' ``like,'' and ``kind of'') from the interview quotes to improve readability. Our approach focuses on analyzing topic summaries \cite{Braun2021-ui} using a combination of deductive (top-down) and inductive (bottom-up) coding. Guided by our research and interview questions, the qualitative data analysis aimed to generate topic summaries that aid in interpreting quantitative results, such as why participants kept using their phones after stating they were willing to stop.

\section{Results}

We have 5,772 valid ESM reports. On average, each participant made 156 valid reports, ranging from 72 to 294. Besides, participants manually dismissed 591 ESM questions in total, and 1,017 ESM questions were auto-dismissed after being present for 30 seconds. We have data of 31,349 usage sessions, later referred to as the full session data, including both those with and without ESM questions. Unless otherwise stated, the following data analysis is based on the data of 5,772 sessions with ESM questions. All data are available on our OSF project\footnote{\url{https://osf.io/x6yzs/}}.

To answer the three RQs, we first describe the distribution of intention and behavior and calculate the IBG, then explain the effects of demographic and contextual factors on the three variables through linear mixed-effects models, and at last develop machine learning models aiming at predicting the three variables in real time. 

\subsection{Describing the Intention-Behavior Gap}
In RQ1, we asked how to operationalize problematic smartphone usage as the intention-behavior gap. To answer it, we first describe intention and behavior separately and then put them together to calculate IBG.  

\subsubsection{The Intention to Continue}
\begin{figure}
  \centering
  \includegraphics[width=0.6\textwidth]{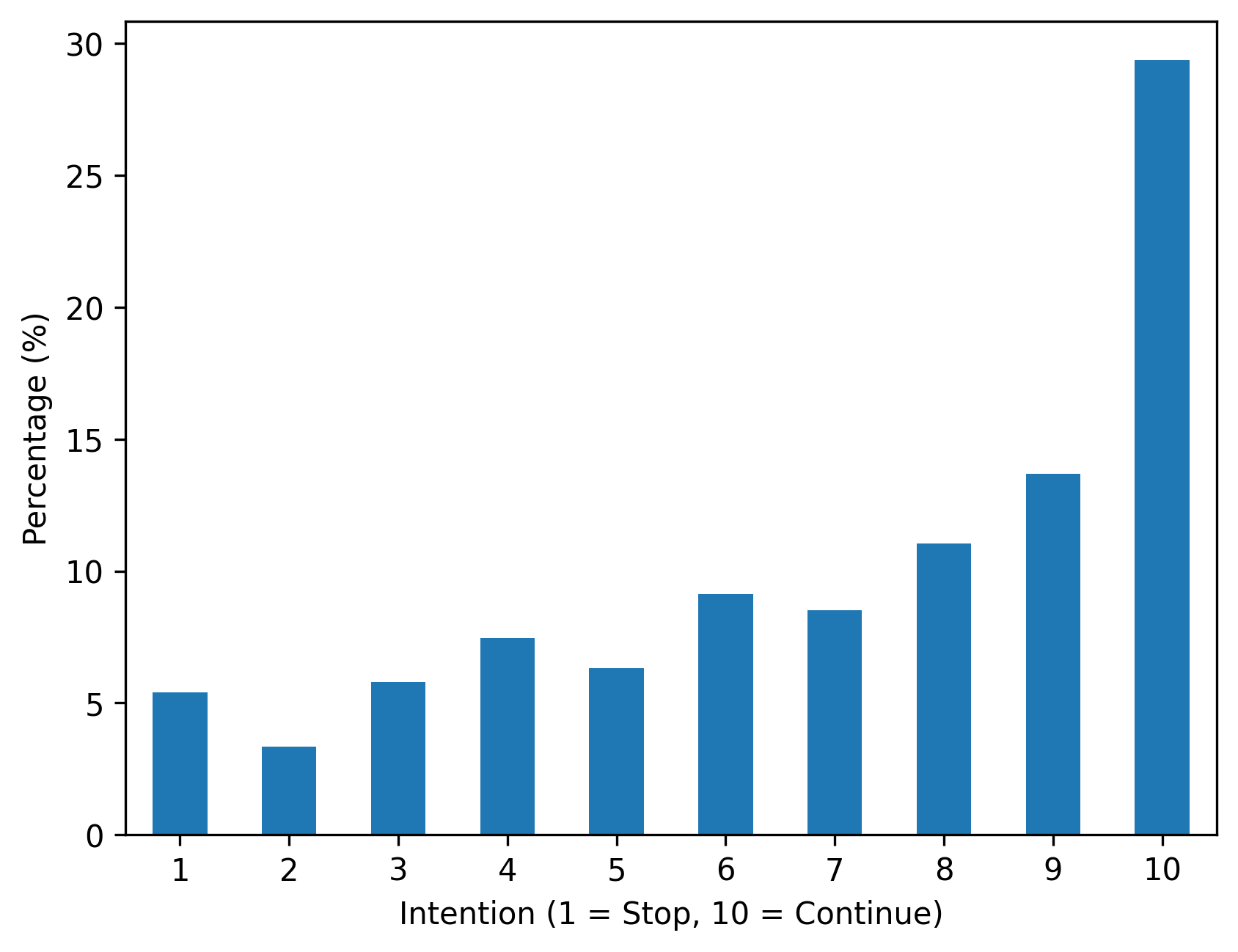}
  \caption{Distribution of the intention to stop or continue smartphone usage.}
  \label{fig:esm_distribution}
  \Description[Bar chart showing response distribution skewed toward continue.]{Bar chart with intention values 1-10 on x-axis and percentage on y-axis. Value 10 (continue) has the highest frequency, followed by 9 and 8. Values 1-5 (stop side) have lower frequencies. The distribution is left-skewed toward continued use.}
\end{figure}

Participants' intention to continue phone usage was heavily left-skewed toward continued use ($M = 7.1, SD = 2.8$, see Fig. \ref{fig:esm_distribution}), with a 25th percentile of 5, a median of 8, and 75\% of responses at the maximum rating of 10. 
Participants would select ``continue'' (8--10) if they were currently engaged in tasks, such as ``reading an article, or browsing through, or doing a survey'' (P04), ``reading something, Instagram, or my Gmail account'' (P06), or ``using my phone to help with an assignment'' (P38). They wanted to ``get it done'' and felt uncomfortable stopping without finishing. 

One of the participants' primary reasons for selecting ``stop'' (i.e., 1--3) was the need to do something physically, such as ``cooking or housework'' (P02), or some tasks that are better carried out on computers, such as attending a meeting. Another common reason for ``stop'' was that participants were aware they should not use their phones for long periods when it was ``close to bedtime'' (P06).

They would select those options in the ``neutral zone'' (4--7) if they were in their free time or ``downtime.'' P15 explained the reason for selecting those numbers: ``I'm just in the neutral zone. I wasn't horrible off that I should just be completely done with it. I could either stop or continue, it wouldn't really matter.'' P08 expressed his ambiguity in that he felt the phone usage was somehow justified but still a waste of time: ``I was probably just sitting on the couch just looking around at stuff. I probably gave it a 5 because there wasn't really any reason that I needed to not be on my phone. It was just kind of my free time, but I felt like I was kind of wasting time.''


\subsubsection{The Behavior of Continuing}
We report the full session length and the remaining session length after the ESM question as the behavior metrics. In later analysis, we focus on the remaining session length. On average, sessions lasted about 17 minutes ($M$ = 1,051 seconds, $SD$ = 1,849 seconds). The median was about 8 minutes (476 seconds), suggesting a right-skewed distribution, with the majority of brief interactions and some extended usage periods. Session length ranged from about 12 seconds to over 11 hours, with 25\% of sessions ending within about 2.5 minutes (160 seconds) and 75\% ending within about 20 minutes (1,208 seconds). 

After answering the ESM question, participants continued to use their phones for about 15 minutes ($M$ = 898 seconds, $SD$ = 1,811 seconds). The median was about 5 minutes (307 seconds), suggesting a similar right-skewed distribution. Remaining session length ranged from about 2 seconds to nearly 11 hours, with 25\% of sessions ending within about 1.5 minutes (89 seconds) and 75\% ending within about 16 minutes (963 seconds).

It is worth noting that those sessions with ESM do not compose a representative sample for all the sessions. Of the full session data, sessions lasted about 8 minutes ($M$ = 479 seconds, $SD$ = 1,561 seconds). The median was about 1.8 minutes (109 seconds), suggesting a highly right-skewed distribution. Session length ranged from less than one second to over 20 hours, with 25\% of sessions ending within about 36 seconds and 75\% ending within about 6.3 minutes (377 seconds). All the results reported here are after screening those long but inactive sessions (see Section 3.4). 

\subsubsection{The Intention-Behavior Gap}
\begin{figure*}
  \centering
  \includegraphics[width=\textwidth]{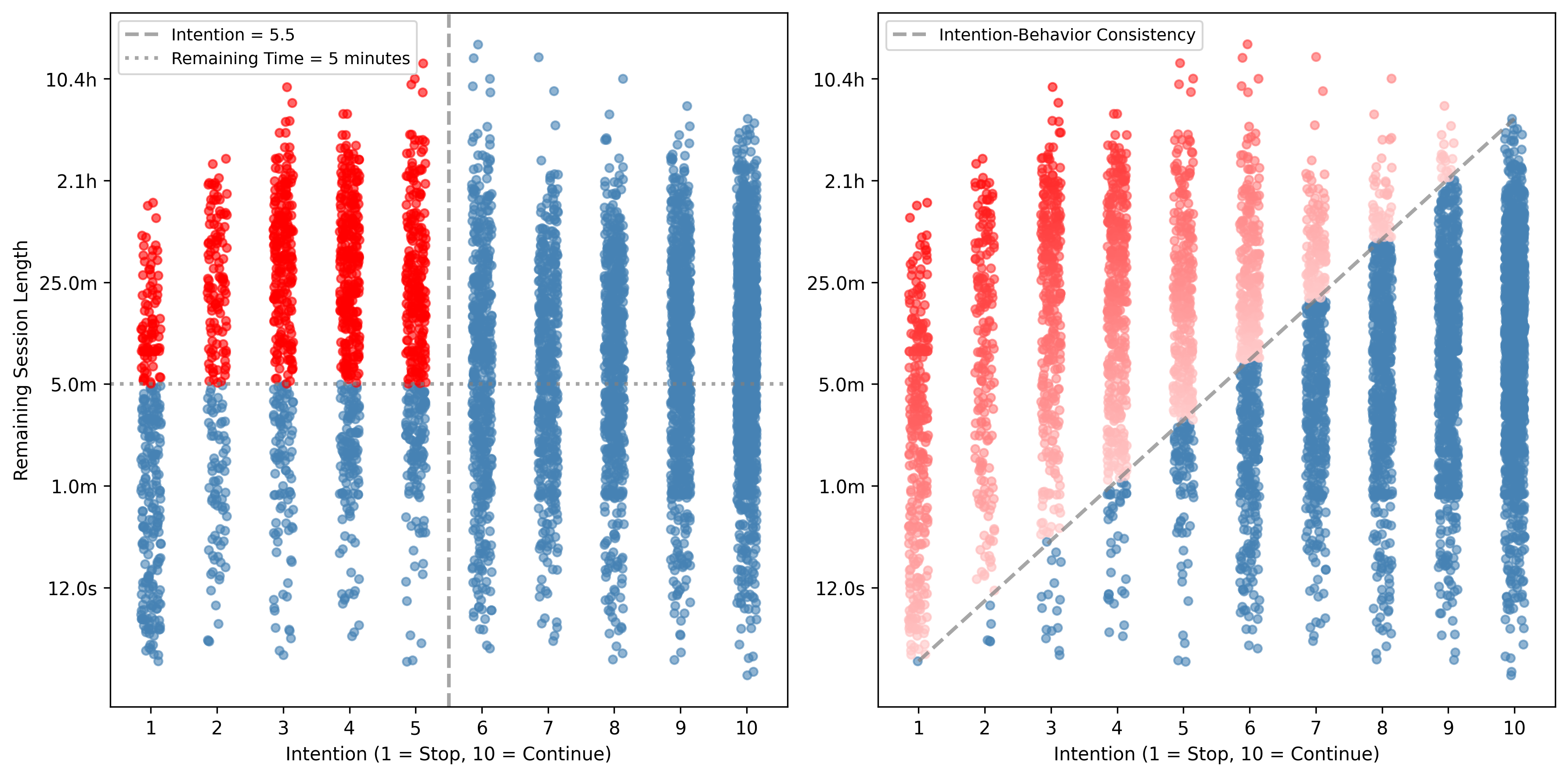}
  \caption{Strip plots showing remaining session length across usage intention. Left: Binary operationalization of problematic usage using arbitrary thresholds---intention below 5.5 (vertical line), remaining length exceeding 5 minutes (horizontal line), and their combination. Right: A continuous operationalization using a diagonal threshold, with red coloring indicating increasing severity of intention-behavior inconsistency on the overuse direction. This comparison demonstrates why a combined measure capturing both intention and behavior can identify problematic smartphone usage more accurately than either dimension alone.}
  \label{fig:time_by_answer}
  \Description[Two strip plots comparing binary and continuous operationalizations of problematic smartphone usage.]{Left: Scatter plot with intention 1-10 on x-axis and remaining time on y-axis. Red points indicate sessions meeting both criteria (intention less than 5.5 and remaining time over 5 minutes); blue otherwise. Right: Same axes with a diagonal line representing intention-behavior consistency. Points above the line are red, with darker red indicating greater inconsistency between intention to stop and actual continued use.}
\end{figure*}

Putting intention and behavior together, we found an overall non-linear relation between them (see Fig. \ref{fig:time_by_answer}). A positive correlation is observed on the left side of the plot (intention values 1--5, $r$ = .30), where higher intention scores correspond to longer remaining session times. However, this relationship diminishes on the right side (intention values 6--10, $r$ = .09), where remaining times are more uniformly distributed across intention levels. Overall, the correlation between intention and remaining time is weak ($r$ = .07).

Fig. \ref{fig:time_by_answer} demonstrates how to define PSU according to the intention and behavior of smartphone usage. If we define and intervene PSU by behavior only (e.g., remaining session length > 5 minutes, 51.4\% of the observations), users will be intervened even if they have justified reasons for the usage and genuinely want to continue. If we define and intervene PSU by intention only (e.g., intention to continue < 5.5, 28.3\% of the observations), users will be intervened in even if they can stop shortly by themselves. Both cases will threaten users' autonomy and cause reactance. By combining intention and behavior (e.g., intention to continue < 5.5 and remaining session length > 5 minutes, 14.9\% of the observations), we can avoid such false alarms, as shown in Fig. \ref{fig:time_by_answer}-left. 

\begin{figure}
  \centering
  \includegraphics[width=0.8\textwidth]{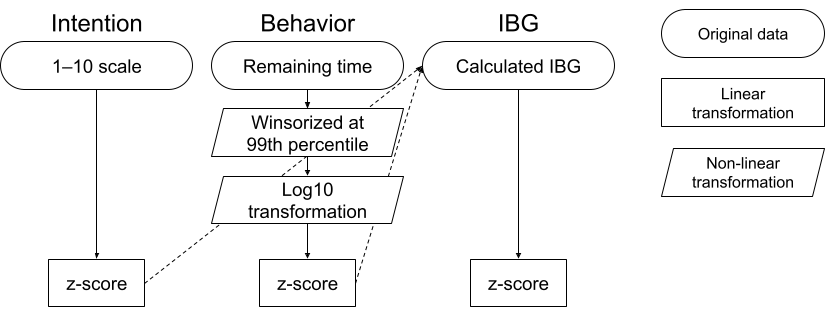}
  \caption{Flowchart of data transformation pipeline.}
  \label{fig:data_pipeline}
  \Description{Flowchart showing how original data have been transformed. Intention, behavior, and intention-behavior gap all start from the original score and end as z-score. }
\end{figure}

\begin{figure}
  \centering
  \includegraphics[width=\textwidth]{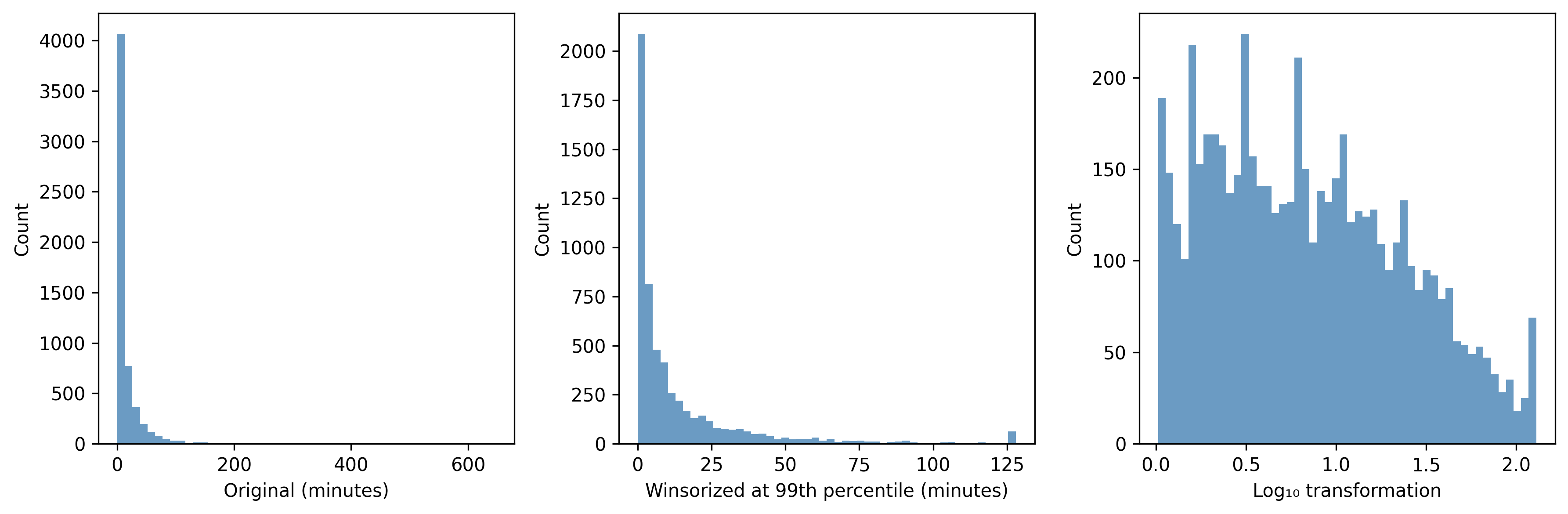}
  \caption{Distribution of remaining session length through the data transformation.}
  \label{fig:remaining_time_distribution}
  \Description{Three histograms showing the distribution of remaining session length. Left: Histogram showing the distribution from 0 to 4000 on the Y axis against 0 to 600 on the X axis; almost all observations are between 0 and 100. Middle: Histogram showing the distribution from 0 to 2000 on the Y axis against from 0 to 125 on the X axis; most observations are between 0 and 75. Right: Histogram showing the distribution from 0 to 200 on the Y axis against 0 to 2 on the X axis.}
\end{figure}

Smartphone usage does not become problematic from non-problematic abruptly. Instead, it varies along a continuum from problematic to justified, with a ``neutral zone'' in the middle. To calculate a continuous IBG, we Winsorized \cite{Beaumont2009-wp} the remaining time at the 99th percentile to limit outliers and applied a $\log_{10}$ transformation to compress the skewed distribution.
To ensure that intention and behavior contribute equally to the gap, we standardized both variables before calculating their difference.
See Fig. \ref{fig:data_pipeline} for the full data pipeline and Fig. \ref{fig:remaining_time_distribution} for the distribution of remaining time through the pipeline.
The comparison between gap values is more meaningful than the values themselves. 
In this initial exploration of IBG in smartphone usage, we chose the most straightforward way of calculating IBG. We don't claim this to be the only or best way of calculation.

There are two types of IBG here: the overuse cases, where participants say they intend to stop but actually continue, and the falling-short cases, where they say they intend to continue but stop shortly. We are more interested in PSU, so we focus on the overuse cases. Nevertheless, we also have some interesting findings on falling-short gaps that we report here briefly. 
We found two main causes for the falling-short gaps. 
The first cause is the completion of the task at hand. Participants' ``continue'' is typically task-bound, not time-bound. 
Saying they would like to stop usually means they want to stop in a short time. Saying they would like to continue, however, does not mean they want to use the phone for long but means they want to finish what they are working on. For example, when asked about a falling-short case, P12 explained ``I did continue for a minute and a half. I know it wasn't a long time, but I completed what I wanted to.''
The second cause is interruption from external factors, such as someone coming or the participant running into someone, so they had to stop. Someone could be family members, co-workers, friends, acquaintances, or even their pets. For example, P04's reason for stopping shortly was ``I was in my email, and then my dog had to be let out, so I had to stop.'' 

\subsection{Explaining the Intention-Behavior Gap}

In RQ2, we asked what factors, both demographic and contextual, could lead to larger intention-behavior gaps in smartphone usage.
To answer it, we fit three linear mixed-effects models to examine predictors of intention, behavior, and IBG. 

\subsubsection{The Linear Mixed-Effects Models}
The participant ID was included as a random intercept to account for the nested structure of repeated observations within individuals ($N$ = 5,772 observations from 37 participants). Fixed effects include demographic variables, app genre at the time of ESM prompt, time period when the ESM was triggered, weekend vs. weekdays, session duration, input interactions in the past minute, and physical movement status.

We used standardized scores of intention, behavior, and IBG as the dependent variables so that the coefficients are more comparable. Table \ref{tab:descriptives} shows the descriptive statistics of the variables, and Table \ref{tab:mixed_models} summarizes the results of the linear mixed-effects models. 
The intention model showed moderate explanatory power (conditional $R^2$ = .583, marginal $R^2$ = .354). The behavior model showed lower explanatory power (conditional $R^2$ = .401, marginal $R^2$ = .263). The intention-behavior gap model performed similarly to the intention model (conditional $R^2$ = .535, marginal $R^2$ = .325). ICC values indicated substantial between-participant variability for intention (.355) and gap (.312) but less for behavior (.187), suggesting a relatively large variance across individuals in the intention.

\begin{table*}[htbp]
\centering
\caption{Descriptive Statistics for Intention-Behavior Gap, Intention, and Behavior and  by Predictor Variables}
\label{tab:descriptives}
\sisetup{table-format=2.2, table-number-alignment=center}
\resizebox{\textwidth}{!}{%
\begin{tabular}{l
                S[table-format=4.0]
                S[table-format=-1.2]
                S[table-format=1.2]
                S[table-format=1.1]
                S[table-format=1.1]
                S[table-format=2.1]
                S[table-format=2.1]}
\toprule
& & \multicolumn{2}{c}{Intention-behavior gap (Z-scored)} & \multicolumn{2}{c}{Intention to continue (1--10)} & \multicolumn{2}{c}{Behavior of continuing (in minute)} \\
\cmidrule(lr){3-4} \cmidrule(lr){5-6} \cmidrule(lr){7-8}
Variable & {\textit{n}} & {\textit{M}} & {\textit{SD}} & {\textit{M}} & {\textit{SD}} & {\textit{M}} & {\textit{SD}} \\
\midrule
\textit{Gender} & & & & & & & \\
\quad Female & 22 & 0.21 & 1.01 & 6.8 & 2.8 & 17.6 & 25.4 \\
\quad Male & 15 & -0.30 & 0.90 & 7.6 & 2.7 & 9.0 & 15.8 \\
\textit{Employment} & & & & & & & \\
\quad Not full-time & 18 & 0.18 & 1.08 & 6.1 & 3.0 & 12.9 & 21.1 \\
\quad Full-time & 19 & -0.18 & 0.88 & 8.1 & 2.3 & 15.3 & 23.5 \\
\textit{Time of Day} & & & & & & & \\
\quad Morning & 1898 & -0.03 & 0.97 & 7.0 & 2.9 & 13.1 & 21.4 \\
\quad Afternoon & 2243 & -0.09 & 0.99 & 7.3 & 2.7 & 13.2 & 21.7 \\
\quad Evening & 987 & 0.03 & 0.97 & 7.3 & 2.8 & 15.3 & 22.8 \\
\quad Night & 644 & 0.35 & 1.09 & 6.3 & 2.9 & 18.3 & 25.8 \\
\textit{Weekend} & & & & & & & \\
\quad No & 4176 & -0.02 & 1.00 & 7.1 & 2.8 & 13.8 & 22.5 \\
\quad Yes & 1596 & 0.05 & 0.99 & 7.1 & 2.8 & 14.8 & 22.1 \\
\textit{App Category} & & & & & & & \\
\quad Message & 2684 & -0.10 & 0.98 & 7.0 & 2.9 & 11.6 & 20.1 \\
\quad Social & 908 & 0.35 & 1.02 & 6.6 & 2.8 & 19.7 & 26.5 \\
\quad Work \& Study & 676 & -0.10 & 0.95 & 7.6 & 2.6 & 15.1 & 23.9 \\
\quad Browser & 470 & 0.03 & 0.97 & 7.3 & 2.7 & 14.1 & 19.3 \\
\quad Daily Life & 408 & -0.08 & 1.02 & 7.2 & 3.1 & 12.1 & 19.5 \\
\quad Media & 396 & 0.15 & 1.03 & 7.0 & 2.9 & 16.9 & 25.3 \\
\quad System Tool & 137 & -0.30 & 0.91 & 8.0 & 2.2 & 14.6 & 27.5 \\
\quad Game & 93 & 0.11 & 0.90 & 7.8 & 2.5 & 18.6 & 22.9 \\
\textit{Scroll Intensity} & & & & & & & \\
\quad Low & 3783 & -0.07 & 0.98 & 7.2 & 2.8 & 13.0 & 21.7 \\
\quad Medium & 1009 & 0.01 & 0.96 & 7.2 & 2.8 & 15.0 & 22.9 \\
\quad High & 980 & 0.28 & 1.08 & 6.6 & 3.0 & 17.3 & 24.0 \\
\textit{Tap Intensity} & & & & & & & \\
\quad Low & 3235 & -0.00 & 1.02 & 7.1 & 2.8 & 14.2 & 22.7 \\
\quad Medium & 1337 & -0.05 & 0.96 & 7.1 & 2.9 & 12.8 & 20.6 \\
\quad High & 1200 & 0.06 & 1.00 & 7.0 & 2.9 & 15.2 & 23.3 \\
\textit{Text Edit Intensity} & & & & & & & \\
\quad Low & 4891 & -0.00 & 1.00 & 7.1 & 2.8 & 14.1 & 22.2 \\
\quad Medium & 422 & 0.12 & 1.02 & 7.0 & 2.9 & 17.1 & 25.3 \\
\quad High & 459 & -0.06 & 0.98 & 6.9 & 3.0 & 11.6 & 21.2 \\
\textit{Physical Activity} & & & & & & & \\
\quad Still & 5315 & 0.03 & 1.01 & 7.1 & 2.9 & 14.7 & 22.9 \\
\quad Moving & 457 & -0.29 & 0.81 & 7.2 & 2.5 & 7.5 & 13.6 \\
\textit{Session Duration} & & & & & & & \\
\quad 0–5 min & 4761 & -0.05 & 0.99 & 7.1 & 2.8 & 13.0 & 21.5 \\
\quad 5–10 min & 683 & 0.19 & 0.99 & 7.2 & 2.8 & 18.1 & 24.5 \\
\quad 10+ min & 328 & 0.34 & 0.99 & 7.2 & 2.9 & 22.2 & 27.2 \\
\bottomrule
\addlinespace
\multicolumn{8}{l}{\footnotesize \textit{Note}. For Gender and Employment, \textit{n} represents the number of participants; for all other variables, \textit{n} represents the number of observations.}
\end{tabular}%
}
\end{table*}

\begin{table*}[htbp]
\centering
\caption{Mixed-Effects Model Results for Intention-Behavior Gap, Intention, and Behavior (all Z-Scored)}
\label{tab:mixed_models}
\resizebox{\textwidth}{!}{%
\begin{tabular}{lrrrrrrrrrrrr}
\toprule
& \multicolumn{4}{c}{Intention-behavior Gap} & \multicolumn{4}{c}{Intention to continue} & \multicolumn{4}{c}{Behavior of continuing} \\
\cmidrule(lr){2-5} \cmidrule(lr){6-9} \cmidrule(lr){10-13}
Variable & $\beta$ & \textit{SE} & \textit{p} & 95\% CI & $\beta$ & \textit{SE} & \textit{p} & 95\% CI & $\beta$ & \textit{SE} & \textit{p} & 95\% CI \\
\midrule
(Intercept) & 0.10 & 0.31 & .734 & [--0.49, 0.70] & \textbf{--0.87} & 0.32 & .007 & [--1.50, --0.24] & \textbf{--0.73} & 0.24 & .002 & [--1.19, --0.26] \\
\addlinespace
\multicolumn{13}{l}{\textit{Demographics}} \\
Gender (male) & \textbf{--0.49} & 0.18 & .005 & [--0.84, --0.15] & 0.35 & 0.19 & .059 & [--0.01, 0.71] & \textbf{--0.34} & 0.14 & .014 & [--0.61, --0.07] \\
Age & 0.00 & 0.01 & .701 & [--0.01, 0.02] & 0.01 & 0.01 & .143 & [--0.00, 0.02] & \textbf{0.01} & 0.01 & .007 & [0.00, 0.02] \\
Full-time employed & --0.25 & 0.18 & .161 & [--0.59, 0.10] & \textbf{0.51} & 0.19 & .006 & [0.15, 0.88] & 0.17 & 0.14 & .230 & [--0.10, 0.44] \\
\addlinespace
\multicolumn{13}{l}{\textit{Temporal features}} \\
Afternoon (ref: morning) & \textbf{--0.07} & 0.02 & .007 & [--0.11, --0.02] & 0.04 & 0.02 & .078 & [--0.01, 0.09] & --0.05 & 0.03 & .064 & [--0.10, 0.00] \\
Evening (ref: morning) & 0.02 & 0.03 & .523 & [--0.04, 0.08] & 0.00 & 0.03 & .978 & [--0.06, 0.06] & 0.03 & 0.03 & .379 & [--0.04, 0.10] \\
Night (ref: morning) & \textbf{0.32} & 0.04 & <.001 & [0.25, 0.39] & \textbf{--0.39} & 0.04 & <.001 & [--0.46, --0.32] & 0.05 & 0.04 & .187 & [--0.03, 0.13] \\
Weekend & 0.02 & 0.02 & .387 & [--0.03, 0.07] & 0.01 & 0.02 & .821 & [--0.04, 0.05] & 0.03 & 0.03 & .190 & [--0.02, 0.08] \\
\addlinespace
\multicolumn{13}{l}{\textit{Category of the current in-usage application (ref: Message)}} \\
Browser & \textbf{0.08} & 0.04 & .049 & [0.00, 0.16] & \textbf{0.09} & 0.04 & .019 & [0.02, 0.17] & \textbf{0.20} & 0.04 & <.001 & [0.12, 0.29] \\
Daily Life & 0.00 & 0.04 & .975 & [--0.08, 0.08] & \textbf{0.09} & 0.04 & .029 & [0.01, 0.17] & 0.09 & 0.05 & .050 & [0.00, 0.18] \\
Game & \textbf{0.22} & 0.09 & .010 & [0.05, 0.39] & 0.04 & 0.08 & .656 & [--0.13, 0.20] & \textbf{0.35} & 0.09 & <.001 & [0.16, 0.53] \\
Media & \textbf{0.31} & 0.04 & <.001 & [0.22, 0.39] & --0.02 & 0.04 & .564 & [--0.11, 0.06] & \textbf{0.40} & 0.05 & <.001 & [0.31, 0.49] \\
Social & \textbf{0.14} & 0.03 & <.001 & [0.08, 0.21] & 0.06 & 0.03 & .072 & [--0.01, 0.12] & \textbf{0.26} & 0.04 & <.001 & [0.19, 0.33] \\
System Tool & --0.03 & 0.07 & .708 & [--0.16, 0.11] & 0.05 & 0.07 & .487 & [--0.08, 0.18] & 0.01 & 0.08 & .866 & [--0.14, 0.16] \\
Work \& Study & 0.02 & 0.03 & .514 & [--0.05, 0.09] & \textbf{0.08} & 0.03 & .016 & [0.02, 0.14] & \textbf{0.11} & 0.04 & .003 & [0.04, 0.18] \\
\addlinespace
\multicolumn{13}{l}{\textit{Input interactions in the last minute (ref: low)}} \\
Scroll mid & 0.04 & 0.03 & .139 & [--0.01, 0.10] & 0.05 & 0.03 & .056 & [--0.00, 0.11] & \textbf{0.11} & 0.03 & <.001 & [0.05, 0.17] \\
Scroll high & \textbf{0.12} & 0.03 & <.001 & [0.06, 0.18] & \textbf{--0.09} & 0.03 & .005 & [--0.15, --0.03] & \textbf{0.08} & 0.03 & .024 & [0.01, 0.14] \\
Tap mid & \textbf{--0.08} & 0.03 & .002 & [--0.14, --0.03] & \textbf{0.07} & 0.03 & .007 & [0.02, 0.12] & --0.05 & 0.03 & .104 & [--0.10, 0.01] \\
Tap high & --0.03 & 0.03 & .364 & [--0.09, 0.03] & \textbf{0.06} & 0.03 & .039 & [0.00, 0.12] & 0.02 & 0.03 & .538 & [--0.04, 0.08] \\
Text-edit mid & --0.01 & 0.04 & .736 & [--0.10, 0.07] & 0.05 & 0.04 & .260 & [--0.03, 0.13] & 0.03 & 0.05 & .540 & [--0.06, 0.12] \\
Text-edit high & \textbf{--0.08} & 0.04 & .041 & [--0.16, --0.00] & \textbf{0.11} & 0.04 & .003 & [0.04, 0.19] & --0.00 & 0.04 & .990 & [--0.08, 0.08] \\
\addlinespace
\multicolumn{13}{l}{\textit{Physical activities}} \\
Moving (ref: still) & \textbf{--0.23} & 0.04 & <.001 & [--0.31, --0.15] & \textbf{0.14} & 0.04 & <.001 & [0.06, 0.21] & \textbf{--0.18} & 0.04 & <.001 & [--0.27, --0.10] \\
\addlinespace
\multicolumn{13}{l}{\textit{Session length so far till the ESM question (ref: <5 min)}} \\
5--10 min & \textbf{0.15} & 0.03 & <.001 & [0.09, 0.21] & \textbf{0.07} & 0.03 & .026 & [0.01, 0.13] & \textbf{0.28} & 0.04 & <.001 & [0.21, 0.35] \\
10+ min & \textbf{0.19} & 0.05 & <.001 & [0.10, 0.28] & 0.04 & 0.04 & .336 & [--0.04, 0.13] & \textbf{0.31} & 0.05 & <.001 & [0.21, 0.41] \\
\bottomrule
\addlinespace
\multicolumn{13}{l}{\footnotesize \textit{Note}. Bold $\beta$ indicates \textit{p} < .05. CI = confidence interval.}
\end{tabular}%
}
\end{table*}

\subsubsection{Effects of Demographic Factors}
Gender (0 = female, 1 = male) had significant effects on both behavior and IBG and a marginally significant effect on intention. Compared with male participants, female participants had larger IBGs, which come from both lower intention and longer remaining session length.
Age had a significant effect on behavior only. Older participants showed longer remaining session length.
Employment (0 = not full-time employed, 1 = full-time employed) had a significant effect on intention only. Full-time employed participants reported higher intention to continue compared with other participants.
Given the relatively small sample size ($N$ = 37), findings regarding demographic effects should be interpreted with caution.

\subsubsection{Effects of Temporal Factors}
\begin{figure}
  \centering
  \includegraphics[width=0.6\textwidth]{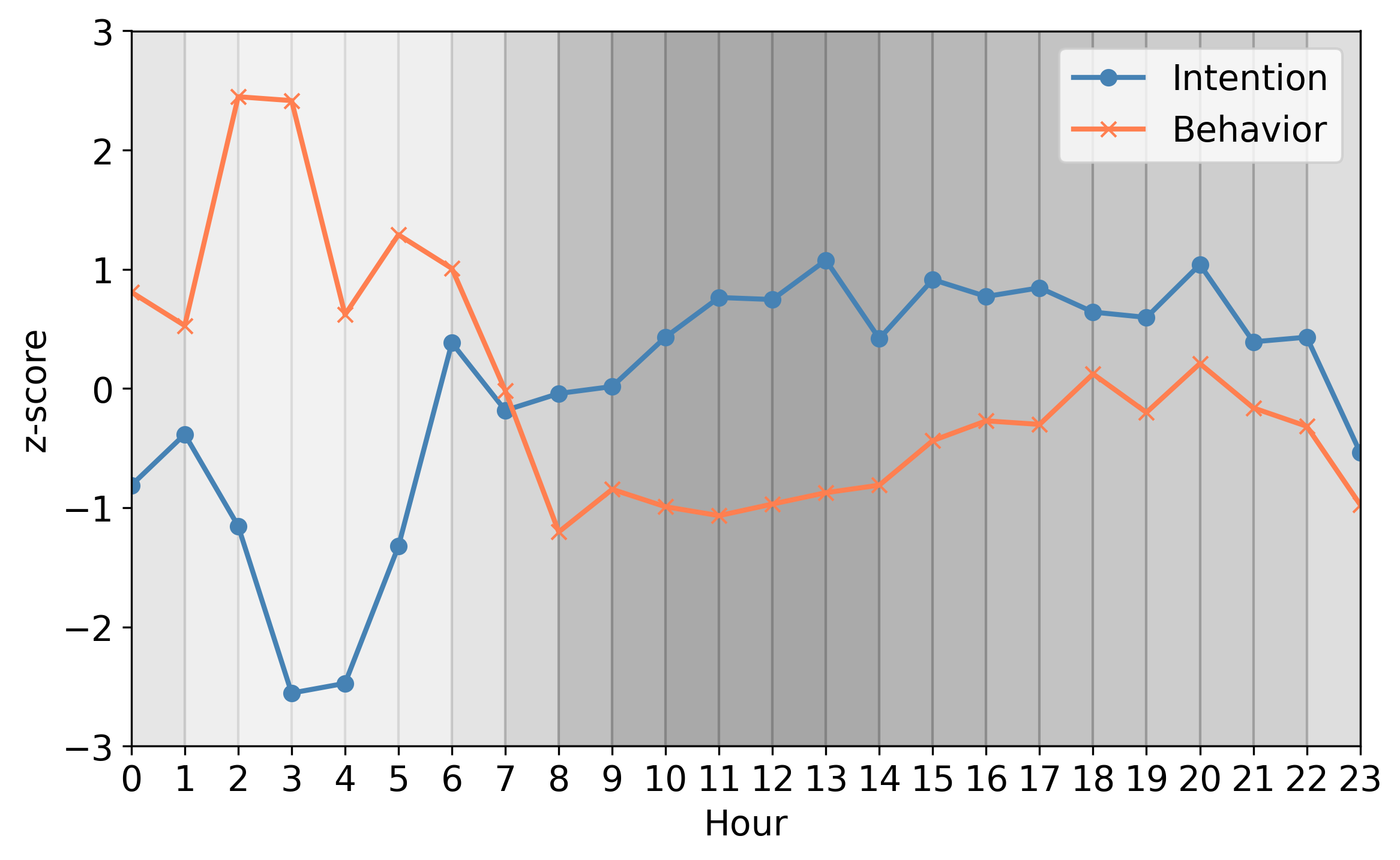}
  \caption{Intention-behavior gap through the 24 hours. Fill opacity is proportional to observation frequency at each hour (darker = more observations).}
  \label{fig:gap_by_hour}
  \Description{Line graph showing standardized intention and behavior from -3 to 3 on the Y axis against the hour from 0 to 23 on the X axis. The behavior line is above the intention line from 0 to 7 o'clock, and below it otherwise.}
\end{figure}

Time of the day has a significant effect on IBG. Through data exploration, we learned that the relationship between hour (0--23) and the IBG was not linear and thus converted hour into four time periods: morning (6:00--11:59), afternoon (12:00--17:59), evening (18:00--21:59), and night (22:00--5:59). 
Compared with morning, nighttime was associated with larger IBGs, which were caused by the lower intentions and the longer remaining session length, as shown in Fig. \ref{fig:gap_by_hour}. 
From the interviews, we saw a recurring theme of PSU associated with sleep. Participants had the habit of using phones when they were ``in and out of sleep'' (P08), ``halfway falling asleep'' (P10), or ``having a really hard time sleeping'' (P34). Some hoped using phones could make them tired enough. For example, P37 used his phone to relieve his anxiety of insomnia:
\begin{quote}
    I would really have a lot of insomnia. [...] and I would really have a lot of anxiety over sleeping. [...] I would try to not use (my phone), like put my phone in another room, on another floor, try to lock it, or something. I had mixed results, but most of the time, it increased my sleep anxiety, because I was putting so much attention towards wanting to sleep. So, now, even though I know it's not ideal, I still use my phone for the sake of just relaxing and managing that sleep anxiety. Which I'm aware is not ideal, but it's good enough for now.
\end{quote}
Similar gaps were also observed when participants just woke up. Participants had the habit of checking news or notifications after waking up but sometimes got distracted and failed to stop, partly because of the lack of willpower when they were still in bed and not fully awake.  

\subsubsection{Effects of the App Category}
To analyze the effect of the app in current usage. We categorized 1,692 apps from our dataset into nine functional categories. App genre information was first retrieved from Google Play, and the original 43 genres were then consolidated into nine categories through manual, function-based reclassification.
To facilitate the manual categorization, we built a web application\footnote{\url{https://app-categorization.vercel.app/}} where we can drag and drop the apps to different levels of categories. The full categorization of the 1,692 apps could be found on OSF \footnote{\url{https://osf.io/x6yzs/files/gx8km}}.

App category affected IBG. Compared with message apps, participants had larger IBG when using browser, game, social, and media apps. For example, P31 explained her reasons for choosing to stop or continue:
\begin{quote}
    If I was doing something like schoolwork, I would say, no, I still need to continue using my phone. But if I was doing something like watching videos on TikTok, and it was just for entertainment, and it was nothing like education, or nothing that was productive, then that's when I would say, okay, maybe I should stop.
\end{quote}
However, usage of those apps is not always associated with PSU and needs to be stopped. P04 explained that ``I've already been here (working) for about an hour and a half, so it's probably time for me to take a little break.'' Similarly, P03 felt she was entertaining herself and ``having a good time,'' and P10 was ``intentionally relaxing'' and ``did not feel like it's a problem.''




\subsubsection{Effects of the Passed Session Length}
Session length before the ESM question also affected the IBG. It is counterintuitive that the longer participants have been using their phones before the ESM questions, the longer they were going to continue using the phones after the ESM questions. This pattern may reflect the uneven distribution of session length---a majority of brief sessions and occasional excessively long sessions. 
From the interviews, many participants said the session length so far was the primary factor they would consider when deciding whether to stop or continue. If they felt they had been on their phones too long, they would select stop. Otherwise, they would continue. For example, P08 said, ``If I just pick up my phone and unlock it. And this pops up right away. I'd be like, no, I don't need to stop using it, because I'm just starting using it.''
However, we did not see this pattern in the quantitative data. One possible reason is that participants could not estimate the session length accurately. They might underestimate for how long they have been using their phones and thus did not see the need to stop soon.
P12 was surprised seeing how long she actually used the phone: ``I can't believe I used a half hour longer than what I intended. I think when you see the visualization, it's like, oh, shit, like, crap, I definitely used it for a long time.''

\subsubsection{Effects of Behavioral Factors}
Input interaction in the last minute had effects on the IBG. For scroll distance, tap count, and text edit count, we excluded zero values and computed the 33rd and 67th percentiles from the non-zero distribution. We then created categorical variables (low, mid, and high) based on these thresholds, with zero and below-33rd-percentile serving as the reference. High intensity of scrolling was associated with larger IBGs. High intensity of text editing was associated with smaller IBGs. Participants did not mention input interactions in the interviews.

To analyze the effect of physical activity, we combined walking, running, and cycling into ``moving,'' because of the few observations of each case. Compared with staying still, moving was associated with shorter remaining session length and smaller gaps. One possible reason is that participants were aware that using phones for long while moving was not safe, as reflected in the effect on intention. Another reason is interruption from external factors, as reflected in the effect on behavior. For example, P40 lived on campus. She ``usually run into people'' she knows and has to stop the phone usage. 

\subsection{Predicting the Intention-Behavior Gap}
In RQ3, we asked how to predict the smartphone usage intention-behavior gap in real time.
To answer it, we explore predicting IBG in real-time. 
If we can predict IBG in real-time, we can build interventions using the predicted IBG as the trigger. 

\subsubsection{Model Design}
We chose random forest regression as the algorithm for the following reasons: random forest regression can handle both continuous and categorical data; it has been used by previous smartphone usage modeling \cite{Orzikulova2024-yd, Pillai2023-dl, Schoedel2023-yv}; our relatively small dataset does not support more complex models like neural networks. We followed the previous study \cite{Orzikulova2024-yd} in the model parameter settings (i.e., number of estimators: 100, max depth: 10, min samples split: 5).
We included 35 features across the following categories: temporal features, interaction, app category and usage metrics, physical activity, cumulative screen time and historical usage metrics of the day, and demographic variables. 

We compared three modeling approaches in predicting the target variables: global, personal, and combined models.
The \textit{global model}, implemented using leave-one-participant-out cross-validation, trains on data from all other participants to predict a held-out participant's outcomes. This approach requires no personal data at deployment, making it suitable for new users. However, it performs well only when the relationship between features and outcomes is consistent across individuals. The $R^2$ is calculated against the global baseline models, which use the mean value of the training data of all the other participants to predict the ``new'' participant's usage. 

The \textit{personal model} trains exclusively on the individual's own data, without incorporating data from other participants. This approach captures idiosyncratic patterns but requires sufficient personal data before making reliable predictions, leading to a cold-start problem. To address this, we began predictions after the first week of data collection and updated the model daily thereafter. Specifically, we used data from Days 1--7 to predict Day 8, Days 1--8 to predict Day 9, and so on. The $R^2$ is calculated against the personal baseline models, which use the mean value of the training data from previous days to predict that of the next day of the same participant. 

The \textit{combined model} integrates data from all other participants with a subset of the target individual's data. Following the same temporal structure, we began predictions after week 1 and updated the model daily. We combined other participants' data with the individual's accumulated data, applying a 10x weight to personal observations. It resembles fine-tuning, but random forest does not support real fine-tuning. It is retrained each day instead. The $R^2$ is calculated against the same personal baseline models.
The combined model is more complex, but does not necessarily outperform the simpler approaches. It outperforms both only when global and personal data provide complementary signals. However, it may perform worse than the personal model when global data introduces noise to individually-driven patterns, and worse than the global model when personal data is too limited or noisy to improve upon population-level patterns.

\subsubsection{Model Performance}
Table \ref{tab:model_comparison} summarizes the model performance. 
For intention prediction, the personal daily model had the best performance. It trains exclusively on the user's accumulated data, without incorporating global data. Adding global data (i.e., the combined model) did not improve the performance. The difference between the performance of the three models suggests contextual effects on intention vary substantially across individuals. The most important features include the average length of previous sessions on that day, the length of this session so far, the weekday, the count of app switches on that day, and the daily accumulative usage length of the app in use right before the ESM question. A mean absolute error (MAE) of 0.528 in z-score of intention corresponds to 1.514 on a 1--10 scale.

For behavior prediction, the combined model had the best performance. It combines global data with the user's personal data. Contextual effects on session duration are more consistent across users than those on intention. The most important features include the length of the last session, the participant's age, the average length of previous sessions on that day, the length of this session so far, and the accumulative screen time so far on that day. An MAE of 0.660 in z-score of behavior corresponds to 10.3 minutes.

The combined model also performed best in predicting IBG. The most important features include the length of the last session, the participant's age and education level, the average length of previous sessions on that day, the length of this app episode so far, and the hour of the day. 

\begin{table}[htbp]
\centering
\caption{Model Performance Comparison of Intention, Behavior, and Intention-Behavior Gap (all Z-Scored)}
\label{tab:model_comparison}
\begin{tabular}{lcccccc}
\toprule
& \multicolumn{2}{c}{Intention to continue} & \multicolumn{2}{c}{Behavior of continuing} & \multicolumn{2}{c}{Intention-behavior gap} \\
\cmidrule(lr){2-3} \cmidrule(lr){4-5} \cmidrule(lr){6-7}
Model & MAE & $R^2$ & MAE & $R^2$ & MAE & $R^2$ \\
\midrule
Global baseline & 0.876 & -- & 0.852 & -- & 0.824 & -- \\
Global & 0.882 & -.139 & 0.712 & .247 & 0.794 & .035 \\
Personal baseline & 0.640 & -- & 0.704 & -- & 0.675 & -- \\
Personal & \textbf{0.528} & \textbf{.242} & 0.673 & .068 & 0.612 & .169 \\
Combined & 0.530 & .229 & \textbf{0.660} & \textbf{.107} & \textbf{0.599} & \textbf{.208} \\
\bottomrule
\addlinespace
\multicolumn{7}{l}{\footnotesize \textit{Note}. Bold font indicates the relatively best-performing model.}\\
\multicolumn{7}{l}{\footnotesize \textsuperscript{a}An MAE of 0.528 in the z-score of intention corresponds to 1.514 on a 1--10 scale.}\\
\multicolumn{7}{l}{\footnotesize \textsuperscript{b}An MAE of 0.660 in the z-score of behavior of continuing corresponds to 10.3 minutes.}\\
\end{tabular}
\end{table}

\section{Discussion}
In this discussion, we contextualize our work by linking IBG to self-regulation theories, comparing findings with prior studies, and exploring the implications of measuring PSU using ESM. We further elaborate on how operationalizing PSU as IBG can inform the design of future interventions. Finally, we address the study's limitations and propose directions for future research. 


\subsection{Connecting Intention-Behavior Gap to Theories of Self-Regulation}


Research on self-regulation and behavior change has examined how to encourage or discourage target behaviors. While this work focuses on the output behavior, it also considers intention as one influential factor, alongside concepts such as ability, motivation, and goals. Research on IBG, in contrast, focuses on the relationship, especially the discrepancy, between intention and behavior. To our knowledge, prior IBG research has not explored smartphone usage, and they usually investigate individual differences over periods of weeks or months. In our study, we explore IBG at a much finer granularity: within usage sessions.

Although self-regulation theories do not directly address IBG, they provide insights into why such gaps occur and how they might be addressed. Dual system theory \cite{Kahneman2012-mz, Strack2004-pf} posits that behavior is determined by two competing systems: the non-conscious System 1 and the conscious System 2 \cite{Lyngs2019-qv}. For a deliberate intention to be translated into behavior, System 2 needs to be activated. Otherwise, System 1 dominates, and people act on habit. In smartphone usage, a lack of conscious monitoring through System 2 often leads to habitual actions driven by System 1, such as browsing on social media aimlessly \cite{Oulasvirta2012-en}. Consistent with this view, we observed larger IBGs while participants were using social media and playing games. Interventions raising awareness and encouraging reflection \cite[e.g.,][]{Hiniker2016-sd, Zheng2025-jx} may help activate System 2 and reduce IBGs. The Fogg behavior model \cite{Fogg2009-dw} offers a complementary perspective, arguing that three elements are necessary for a behavior to occur: motivation, ability, and a prompt. When motivation and ability are both high but proper prompts are absent, the behavior would not happen. In such cases, a simple signal, such as a reminder, may be enough to trigger the behavior.

Goal system theory \cite{Kruglanski2002-zf} suggests that people have dynamic systems of sometimes competing goals. Temporal self-regulation theory \cite{Hall2007-yh} similarly highlights conflicts between short-term rewards and long-term goals. From this perspective, avoiding excessive smartphone usage---which promotes long-term health---is just one goal embedded within a broader goal system, where it competes with other short-term goals more directly associated with reward. If the intention to put the phone down cannot overcome short-term rewards of novelty-seeking behavior \cite{Kheradmand2023-nh}, it fails to translate into action. Some interventions attempt to strengthen the self-control goal by linking smartphone usage control to financial rewards \cite{Jang2025-ui, Park2021-dw}, though their effectiveness often diminishes once the incentives are withdrawn.

\subsection{Comparison with Previous Studies}
We found that gender, temporal factors, app category, and input interactions can explain the variance in IBG. The results are generally consistent with previous research on PSU, providing support for the validity of conceptualizing PSU as IBG.
Female participants exhibited larger IBG, consistent with previous studies showing that females are more prone to PSU \cite{Busch2021-fi}. We have observed higher IBG when participants were using social media, media entertainment, games, and browsers. These app categories are often perceived as more meaningless \cite{Lukoff2018-ws} and ritualistic \cite{Hiniker2016-sd}.
Although the correlation between poor sleep quality and PSU has been well documented \cite{Busch2021-fi}, sleep problems are usually considered a consequence rather than a cause of PSU. However, based on our interviews, sleep problems may also contribute to PSU. Specifically, attempts to avoid phone use before sleep can induce anxiety, thereby exacerbating insomnia and potentially increasing PSU. Moreover, we have observed a general pattern of larger IBG around sleep hours. This could be explained by the depletion of willpower \cite{Baumeister2007-on, Baumeister2024-ah} at the end of the day. Some cases of PSU immediately after waking up may be linked with the low level of blood glucose, a physical resource of willpower \cite{Gailliot2007-my}.

Previous studies have utilized input interactions as features in machine learning models \cite{Chen2023-dc, Orzikulova2024-yd}, proxies for PSU \cite{Monge-Roffarello2023-af}, or direct intervention targets \cite{Lu2024-dc}. Our study extends this line of work by explicitly examining the relationship between input interactions and PSU. We found that PSU is associated with more frequent or longer scrolling behaviors but fewer text edits. This pattern suggests that the total count of input interactions may not be an effective predictor of PSU.

Our machine learning models achieved relatively low $R^2$ values, indicating modest predictive performance. To contextualize these results, we compared our model performance with that of previous studies.
We predicted intention as a continuous variable, whereas prior work has treated intention as a binary variable. When we adopted the same approach by converting intention into a binary label at a cutoff of 5.5, our random forest model achieved an F1 score of .66, closely aligning with the performance reported in \cite[][F1 = .67]{Orzikulova2024-yd}.
To the best of our knowledge, no prior research has attempted to predict full or remaining session length in real time. The most closely related work we identified predicts remaining surgical case duration \cite{Jiao2022-yt}, reporting an MAE of 13.8 minutes. Our model achieved an MAE of 10.3 minutes for predicting remaining session length, which is comparable in magnitude. Additionally, this study is the first to predict IBG, precluding direct comparisons. Overall, these findings indicate that the performance of our models is broadly comparable to results reported in related prior studies.

\subsection{Measurement of Problematic Smartphone Usage with Experience Sampling Method}
HCI researchers usually employ ESM to collect self-reported PSU data at the session level. Because of the high sampling frequency, these measurements are typically limited to a single question. When the goal is to describe or explain PSU, the measurements are often operationalized on continuous scales \cite[e.g.,][]{Lukoff2023-lt, Rixen2023-wg, Terzimehic2023-yi, Lukoff2018-ws}. In contrast, when the goal is prediction, the measurement is usually simplified into a binary variable \cite[e.g.,][]{Orzikulova2024-yd, Hiniker2016-sd, Chen2023-dc}. Predictive performance often appears stronger when predicting binary labels compared to continuous variables. It is much less likely to achieve a perfect prediction of a single data point on a 10-point scale (as used for intention in this study) than on a binary scale. Moreover, it is virtually impossible to make a perfect prediction for real behavioral data, such as session length measured in milliseconds. However, this binary approach sacrifices variance information and can only support all-or-nothing interventions.

A continuum from normal to excessive has been used to conceptualize media self-regulation \cite{LaRose2003-tr} and daily consumption behavior \cite{Grover2011-mm}. Similarly, PSU should exist along a continuum, spanning from clearly justified to clearly problematic use, with intermediate cases such as mildly or moderately problematic use in between. 
If needed, continuous variables can be converted into categorical (including binary) variable by setting thresholds, but categorical variables cannot be converted to continuous.
From the interviews, we also learned that there is a ``neutral zone,'' where participants do not have a clear intention to stop or continue. Forcing participants to choose one side may fail to capture their original intention. In some cases, people simply do not know their attitude and need to infer it by observing their own behavior \cite{Bem1972-dk}. In such situations, participants may select options in the middle of a scale, but this selection is conceptually distinct from the ``neutral zone.''
In the study of instrumental vs. ritualistic phone usage \cite{Hiniker2016-sd}, participants were also given the option ``I don't know why I was using my phone,'' but those responses were not included in the later analysis. Based on this observation, we recommend that future ESM studies on phone usage include an explicit ``I don't know'' response option.

\subsection{Intervention according to the Intention-Behavior Gap in Smartphone Usage}
Operationalizing PSU as IBG can inform the design of better user-aligned intervention systems. 
Our findings suggest two concrete components to improve existing PSU interventions: (1) IBG-based triggering, where interventions are delivered according to predicted IBG values rather than intention or behavior alone, and (2) continuous intensity scaling, where intervention strength varies gradually with the severity of PSU rather than in a binary manner. 
These components map naturally onto the Just-In-Time Adaptive Intervention (JITAI) framework, which emphasizes delivering interventions at the moments when users need them most \cite{Nahum-Shani2016-ei}. JITAIs consist of six core components: distal outcomes, proximal outcomes, tailoring variables, decision points, decision rules, and intervention options \cite{Nahum-Shani2016-ei, Goldstein2017-fq}. Within this framework, we position IBG as the \textit{tailoring variable}, and continuous intensity scaling as a \textit{decision rule}.
Although we believe the two components work best in combination---and that reflection-based interventions are particularly well suited (see Section 5.1)---each can be decoupled and integrated with various intervention formats, including blocking, self-tracking and reflection, and reward-based designs \cite{Lyngs2019-qv, Monge-Roffarello2022-mw}).

\textbf{IBG-based Triggering.}
Interventions triggered by intention or behavior alone risk delivering interventions when they are not needed, potentially provoking psychological reactance. We propose delivering interventions only when an overuse gap exists between intention and behavior---that is, when users wish to stop but are unable to do so independently. At such moments, external support is genuinely needed and thus more likely to be appreciated. 
In practice, IBG-based triggering can be integrated with most existing interventions by replacing the current trigger, if applicable, with predicted IBG, without altering the intensity scale. For example, if an intervention currently displays a blocking page whenever the phone is unlocked, we can change the trigger from screen-unlocking to predicted IBG reaching a threshold.

\textbf{Continuous Intensity Scaling.}
Most existing interventions operate in a binary mode: either fully active or absent. We argue instead for intensity that varies continuously with the severity of PSU. This principle is well-established in other domains. In medicine, treatments are calibrated to symptom severity, following the principle of using the lowest effective dose. 
In mental health care, levels of support range from Level 1 self-management to Level 5 specialist and acute services.\footnote{\url{https://docs.iar-dst.online/en/v1/level-of-care/overview.html}}
Yet PSU intervention research typically applies the strongest available intervention to all detected problematic usage, often to achieve significant results and large effect sizes.
Although the JITAI explicitly calls for adaptive intervention dose \cite{Nahum-Shani2016-ei}, we are unaware of any prior research that aligns multiple levels of intensity to users' needs.
We propose calibrating intervention intensity to the magnitude of IBG. 
Stronger interventions could take various forms, such as more digits to be typed correctly \cite{Kim2019-sk} or longer delay \cite{Haliburton2024-kb, Nakamura2025-bo} before unlocking the phone or opening an app, larger location offset and longer delay of input interaction \cite{Lu2024-dc}, or higher financial incentives for not using the phone \cite{Jang2025-ui}. For example, LocknType \cite{Kim2019-sk} has interventions of two levels of strength (i.e., 10 digits and 30 digits typing tasks), but treated those two as separate conditions and compared their effect. A continuous-scaling design would instead vary the number of digits with IBG, applying lighter friction for small gaps and heavier friction for large ones.

The simplest decision rule mapping IBG to intensity is linear:
\begin{math}
  I = a \cdot G + b
\end{math} where $I$ is intervention intensity, $G$ is IBG, $a$ controls the slope (how sensitive intensity is to IBG), and $b$ is a baseline offset (which also serves as the threshold in a binary schema). Appropriate values of $a$ and $b$ should be calibrated to specific intervention goals and formats, likely through pilot studies.
Given substantial individual variability in smartphone usage, personalizing these coefficients may further improve effectiveness. In development of PSU questionnaires, the value of $b$ is typically determined through receiver operating characteristics analysis and varies across genders and countries \cite{Kwon2013-zm, Lopez-Fernandez2017-cz}. We also recommend allowing fine-tuning the decision rule by the users to support their autonomy \cite{Alberts2024-ht}.

\subsection{Limitations and Future Directions}
The relatively small sample size and self-selected sample limit the generalizability of the results, especially demographic effects on PSU. It also constrains the stability and complexity of both the linear mixed-effects models and the machine learning models. For example, although we observed that the effects of contextual factors on users’ intention may vary across individuals, we were unable to include random slopes in our mixed-effects models. A larger sample size would enable more robust modeling of individual differences and improve generalizability.

We recruited Android users only; the restrictions from the operating system have been documented in literature \cite{Lyngs2022-xg, Kim2017-ae, Monge-Roffarello2022-mw, Zheng2025-jx}. A new challenge we have encountered is about activity recognition. We designed ESM surveys to be suppressed whenever the Google activity recognition API detected the current physical activity as ``in vehicle.'' However, due to the accuracy and latency of the activity recognition, it is still possible that the participants receive the surveys while driving. Besides, we did not find a reliable way to tell apart drivers from passengers when the detected physical activity is ``in vehicle.'' 
We also recruited participants who were interested in reducing excessive phone usage, so our findings may not generalize to users who lack this interest or who do not perceive their phone use as excessive. Within our recruited sample, we further excluded five participants who consistently selected ``continue'' across nearly all surveys, suggesting little or no perceived intention-behavior gap. One possible explanation is that these users were not genuinely engaged with the study and participated primarily for compensation, which would raise concerns about the quality of their data. Alternatively, they may simply not have experienced IBG during the study period---in which case their exclusion further narrows the population to which our findings apply. This sample limitation also points to a more fundamental question about the scope of PSU interventions. If users do not perceive their own phone use as problematic, do they need interventions at all? And who is positioned to decide otherwise? While some interventions are voluntarily adopted by motivated users, designing tools that target people who report no concern with their usage raises ethical considerations around autonomy and paternalism that future work should address explicitly.

The validity of the data is affected by several factors.
During the interviews, participants sometimes had trouble recalling accurately what happened during the data collection. The sooner the interview takes place, the better chance we can reconstruct what happened. But scheduling timely interviews can be challenging, especially if the participant is full-time employed. 
There are some careless responses from the participants. We attempted to detect such responses using response time \cite{Ulitzsch2024-cq}, but this approach performed poorly in our data, likely because each ESM survey consisted of only one question. We will continue to explore methods for detecting careless responses in single-item ESM surveys. 
The single-item measurement may not fully capture the intention variance, but adding even a second question per ESM will almost double the time participants need to spend in the study (they will need to answer over 200 more questions) and probably affect their willingness to participate till the end.
In the interviews, several participants described a question–behavior effect \cite{Sprott2006-ax}: answering the ESM questions prompted them to reflect on their phone usage and (or) may alter their behavior, although the responses were mixed.

We operationalize PSU as IBG to support more accurate and adaptive intervention systems, but we acknowledge that IBG does not perfectly overlap with PSU. Falling-short gaps, where users intend to continue but stop shortly, are not problematic. 
Our IBG framework is built upon the core concept of current session length to explore whether and when users will stop the usage. However, PSU manifests in various forms, and our framework cannot capture all types of PSU. For example, the ``should-not-start'' cases is left out. Future research may explore framing IBG around other concepts, such as the accumulative daily screen time or discrepancies between intended and actual app usage.
PSU could exist even when the screen stays locked \cite{Heitmayer2021-yh}, but we focused the usage between unlocking and locking of the screen. 

In practical interventions, prediction errors are unavoidable but can be mitigated. Reinforcement learning \cite{Sutton2015-qy} could be used to update the model and potentially improve the model performance. Selective prediction based on confidence \cite{El-Yaniv2010-kj} could also be applied to reduce prediction errors. According to our simulation, the error in prediction is positively correlated with the $SD$ of the predictions across all the decision trees in the random forest. The intervention system can withhold from making predictions if the $SD$ exceeds a threshold.
\section{Conclusion}
We demonstrated a novel approach to operationalizing problematic smartphone usage as the intention-behavior gap (IBG). We found that IBG could be explained by users' gender, time of the day, app category, input interactions (i.e., scrolling and text editing, which showed opposite effects), physical activity, and elapsed session length. Intention was most accurately predicted by a personal model, achieving an MAE of 1.514 on a 10-point scale, whereas behavior was best predicted by a combined personal–global model, with an MAE of 10 minutes. IBG was also predicted most accurately by a combined personal–global model.
As a next step, we plan to build an intervention system that adapts the intervention intensity in real time based on the predicted intention-behavior gap in smartphone usage.

\bibliographystyle{ACM-Reference-Format}
\bibliography{lib2}

\appendix
\section{Appendix Section}
\subsection{Interview protocol}

Hi, good morning/afternoon! How are you? Thank you for joining us today. We appreciate your time and participation. Your responses will be kept confidential and anonymous. Any information you provide will be used solely for research purposes.

With your permission, I would like to audio-record this interview to ensure accuracy. May I have your consent to proceed with the recording?

\textbf{\textit{Background Information}}
\begin{enumerate}
   \item Could you please introduce yourself briefly? Like what do you do?
   \item What are your phone usage habits? 
   \begin{enumerate}
       \item What is the usual purpose for using your phone? 
       \item Is the purpose the same during daytime and night? On weekdays vs weekends?
   \end{enumerate}
   \item Is there any aspect of your phone usage that you would like to control/limit/reduce? 
   \item Why do you want to control it/them?
\end{enumerate}

\textbf{\textit{Data Engagement}}

(Show slides: \textbf{Top five apps})

\begin{enumerate}[resume]
   \item Here are five of your most used apps during the data collection. Could you please tell me the main reason or purpose behind the usage of each app?
\end{enumerate}

(Show slides: ESM question with the choices and the \textbf{distribution of responses})

\begin{enumerate}[resume]
   \item During the data collection, you have answered this question many times. Here is the distribution of your responses. What factors or circumstances generally influenced your decision to continue or stop? Or in the middle?
\end{enumerate}

(Walk the participant through the visualization with 1 to 2 examples)

(Show slides: About three instances where they indicated they \textbf{should stop, but continued} using the phone for a while. For each:)
\begin{enumerate}[resume]
   \item Can you recall what happened at this time? 
   \item Why did you continue to use your phone after answering “I should stop”? Any external or internal reasons?
   \item How do you feel about using your phone longer than you intended?
   \item If the same situation happens again, what will you do?
\end{enumerate}

(Show slides: About three instances where they indicated they \textbf{should continue, but stopped} using the phone soon. For each:)
\begin{enumerate}[resume]
   \item Can you recall what happened at this time? 
   \item Why did you stop using your phone after answering “I should continue”? Any external or internal reasons?
   \item How do you feel about using your phone shorter than you intended?
\end{enumerate}

\textbf{\textit{General Experience During the Study}}
\begin{enumerate}[resume]
   \item How was your general experience during the study period?
   \item What did you learn, if any, during the study period?
   \item What did you like and dislike about using AwareMind?
   \item How did AwareMind influence your understanding about phone usage?
   \item Did the action of answering those questions affect your intention and behavior of phone usage?
   \item What other information may help you recall what happened during the study?
\end{enumerate}

\textbf{\textit{Exiting}}

That is the end of the interview. Do you any questions for me? Let me stop the audio recording. 

Thank you again for your time and valuable insights. Have a great day!


\end{document}